\documentclass[12pt, article, letter]{article}
\usepackage{amsmath, amsthm, amssymb,wasysym} 
\usepackage{setspace}
\usepackage{lineno}
\linenumbers

\usepackage{natbib}
\usepackage[compact]{titlesec}
\usepackage{graphicx}
\usepackage{subfigure}
\titlespacing{\section}{0pt}{*0}{*0}
\titlespacing{\subsection}{0pt}{*0}{*0}
\titlespacing{\subsubsection}{0pt}{*0}{*0}
\usepackage[top=1in, bottom=1.2in,left=0.8in,right=0.8in]{geometry} 
\title{Simulated Mima mounds emerge from small interactions}
\author{Chlo\"e Lewis\\University of California, Berkeley; \texttt{chlewis@berkeley.edu}}

\begin{document}
\maketitle

\begin{abstract}
The Mima-mound-and-vernal-pool topography of California is rich in endemic plant and invertebrate species, but we do not know how this unusual environment is created or maintained. Burrowing rodents have been observed to move soil upwards at annual rates sufficient to maintain the mounds despite erosion, but there is no tested explanation of this behavior. We propose that the mounds are an emergent effect of small-scale (10 cm, 1 day) interactions between topography, hydrology, plant growth, and rodent burrowing. A cellular automata simulation of these processes both generates and maintains mound-pool topography with minimal dependence on initial conditions, and can also describe mound morphogenesis on slopes, where observed mound geometry is distinct from that on level ground. 
\end{abstract}

\section*{Introduction}
\label{sec:intro}
Mound-pool topography, or Mima mounds, were common on many landforms in California, Oregon, and Washington before 1850 \citep{smith-verrill1998}. The numerous small vernal pools of this topography are habitat for endemic species of plants and animals \citep{bradford1992biogeography, skinner1994california}. Better understanding of natural mound-pool maintenance should improve the protection and reconstruction of habitat \citep{zedler2009feedbacks}. 

Mound-pool topography formation and maintenance has been ascribed to earthquakes, erosion and deposition by wind or water, relict tree-stumps, or glacier thaw, but these hypotheses do not fit all the similar mound fields  \citep{washburn1988}. The fossorial hypothesis, that burrowing rodents are responsible \citep{turner1896}, has been gaining strength with detailed observations, usually of pocket gopher activity \citep{Cox:1990, horwath-johnson2006, reed2006}. (Many studies of fossorial rodents refer to the loose material removed from their burrows every few days as `mounds'; these piles are from 10 cm to a meter across, and are unconsolidated, unvegetated sediment. This paper refers to these piles as `tailings', and only to the  vegetated, Mima-type mounds, diameters 1 m to 10 m, with developed pedogenic features, as `mounds'.) It is still unclear what causes gophers to maintain mounds in some regions but not in others.  Centering mounds on the nests or favored foods of gophers \citep{dalquest-scheffer1942, horwath-johnson2006} does not explain why mounds only occur in some of the prairies inhabited by \textit{Thomomys} and \textit{Geomys} species. Assuming that mounds exist only where intermound soil is too thin to contain burrows 
 is contrary to observations of fossorial activity in intermound regions during dry periods (pers. obs., Jepson Prairie, Nov. 2011). In many California grasslands fossorial rodents increase erosion and reduce variation in soil thickness \citep{yoo2005a}, particularly on shallow slopes \citep{Reichman-seabloom2002}, which would tend to erase mounds rather than maintaining them.  Moreover, not only is there no evidence that rodents in mound-pool regions have developed mound-building instincts, there is some evidence against it: \cite{patton1990evolutionary} classifies \textit{Thomomys bottae} of the whole North Central Valley from the Delta to Oregon as one genetic group, with no significant genetic heterogeneity between the two populations in the group. This region includes great swathes of surviving and pre-colonization vernal pool and mound-pool habitat and also non-pool habitat \citep{holland1998}.  Pocket gophers in flood-irrigated alfalfa fields live in the raised field edges \citep{pipas2000}, suggesting that the gopher mound-building response is not an immediate reaction to inundation. 

I hypothesize that surface irregularity and annually-increasing inundation are enough to both localize and direct fossorial activity in a way that amplifies the surface irregularity into mound-depression microtopography without any mound-specific instinct or initial condition. Specifically, plant species and growth is organized in space by irregular soil surface, and in time by seasonal inundation \citep{lazar2004,zedler1987ecology}. Fossorial rodents forage on the organized plants. For instance, pocket gophers, a fossorial rodent common to California,  have food preferences \citep{Hungerford:1976,Tilman:1983}, possibly including phenological preferences\citep{Behrend1988feeding}. Pocket gophers reach population density high enough to disturb each point in a natural grassland every 3 to 5 years\citep{hobbsmooney1995} and can individually move 8 to 66 liters of soil a day \citep{andersen1987geomys}.  I hypothesize that rodent attraction to plants at the water margin biases rodent movement downward and therefore biases soil movement upward. For fossorial activity to  maintain mounds against erosion, the net upward movement of soil must be great enough to counterbalance the rapid erosion of rodent tailings as well as the general erosion rate. 

This model  explains why the Mima mounds are associated with seasonally-inundated landscapes \citep{nikiforoff1941hardpan, smith-verrill1998} more parsimoniously than do explanations depending on either universal or adaptive behavior  \citep{dalquest-scheffer1942,mielke1977,Cox:1987}. Behavior common to an entire rodent species, such as nest-building, should tend to create mounds in all landscapes if it does in any landscape. An emergent process of mound creation is also testable, and consistent with claims  of recent mound re-formation \citep{JohnsonGSA2011}.

The main processes that interact in a mound-pool landscape were described independently and combined in a cellular automata (CA) model of landscape development. In a CA, physical space is represented by discrete cells. Each cell stores the values of one or more state variables, so that each state variable is represented discontinuously over the whole space. The laws of the simulated system are formulated in terms of the state variables in each cell and a specified set of its neighbors. Each law may require that any set of the state variables in the current cell and its neighbors be updated based on their joint states. Time is simulated by updating all the cells, checking that conserved values are in fact conserved. 

CA have two particular strengths. First, the representation of processes at local scales is often a good match for the measurements we can make to test the models. Second, CA can demonstrate features that develop not `top-down', from central or controlling agents, but as emergent processes requiring the interactions of all the constituent parts. Many landscape and ecology problems that have resisted top-down explanations are tractable to CA and other emergent methods \citep{Fonstad2006}. 

\section*{Methods}
\label{sec:methods}

\subsection*{Model representation}
In this model, the simulated landscape is broken into square cells representing an area 10 cm on a side. 10 cm is  approximately the size of a large \textit{Thomomys} burrow or a small surface tailings mound of burrow ejecta. Each cell stores four state variables (Table \ref{tab:StateVars}). Four processes were identified that occur in each simulated day: plant growth, rodent activity, soil erosion, and inundation or evaporation. Processes operate on the state variables and interact only through the state variables. 
\begin{table}[l]
  \caption{State variables maintained by each cell of the CA}

\begin{tabular}[l]{p{1.5in}| p{4.5in}}
\hline
Elevation& Soil surface, from upper limit of the developing restrictive layer.\\ \hline
Vegetation density & A general measure of green plant density, from 0 to 1.\\ \hline
Rodent presence & Whether a fossorial rodent is currently in the cell.\\  \hline
Water table & Depth of the water table above the restrictive layer; in a level soil, this is the same for every cell.\\ \hline
\end{tabular}
  \label{tab:StateVars}
\end{table}

\begin{table}[l]
  \caption{Available processes for each of the four main landscape processes modeled}

  \label{tab:Processes}

\begin{tabular}[l]{p{1in}| p{5in}}
\hline
Hydrology& Well-drained soil, in which no perched water develops and all cells are equally saturated during the rainy season. \textit{Fails}\\ \hline
& Restricted-drainage soil, in which the perched water table rises during the rainy season and evaporates away in the dry season.\\ \hline
& Overland flow, in which cells below the water table periodically move in a fixed direction, potentially picking up or depositing loose sediment.\\ \hline 
Plant growth & Plants grow faster when closer to the water level, in a linear relation.\\ \hline
& Plants grow equally well everywhere, during the rainy season. \textit{Fails}\\ \hline
Erosion & Soil moves between Von Neumann-neighbor cells, proportionally to the lack of vegetation and the elevation difference. Fast, and develops mound-pool landscapes when Moore-neighborhood erosion does, but the results show the alignment of the grid (see Fig. \ref{fig:erosion}).\\ \hline
& Soil moves between Moore-neighbor cells, proportionally to the lack of vegetation and the elevation difference.\\ \hline
Rodent activity & Rodents will not move or stay below the water table, and cannot occupy a cell already occupied by another rodent. They will move to the adjoining cell with the greatest vegetation; if neighboring cells are tied, they choose randomly. \textit{Fails}\\ \hline 
& As above, but rodents will not move into a cell adjacent to another rodent. \\ \hline
& Rodents will not move or stay below the water table, and cannot occupy a cell already occupied by another rodent. Among acceptable cells, they change direction with a probability proportional to the richness of the vegetation in their current cell. \textit{Fails}\\ \hline

\end{tabular}
\end{table}

Each process is described with several alternate algorithms representing competing common descriptions of the real-world process (Table \ref{tab:Processes}). A complete simulation of a landscape requires that the initial topography be specified and each of the four processes be assigned one specific algorithm of the several available. The simulation then goes forward by running each of the sub-processes in turn for each simulated day. This system allows simulated comparison and manipulation experiments; one simulation may be copied, and one state variable array or  sub-process in the copy changed to an alternative. The pair of simulations are run in parallel, and the results compared. 

The model is written in Python, using \textit{numpy.random} where stochastic behavior needs to be simulated and \textit{PIL.Image} and \textit{scipy.ndimage} for image analysis \citep{jones2001python}. 

\subsection*{Parametrization}

The model is parameterized with measurements made at Jepson Prairie Nature Reserve, a well-studied, pedologically young \citep{smith-verrill1998} basin-rim formation in the Central Valley of California. Jepson Prairie includes mound-pool formations in the Antioch-San Ysidro Complex. Where these measurements cannot be made, we use values from the literature for the Central Valley. Sensitivity testing is done by comparing output simulated with parameter values at the mean and extrema of given ranges. 

Where measurements are not available, the model uses values or algorithms that are pessimal for the generation of microtopographic structure. For instance, the initial topography of a site could be perfectly flat or randomly irregular on the 10 cm scale. The introduced simulated rodents were generally added at the center of the model, not randomly across the landscape. These were chosen to make mound-pool initiation less likely than  in the real-world cases: for instance, a real California grassland before the development of a restrictive claypan or hardpan might have had grass tussocks or shrubs at meter or ten-meter average spacing, with a slight elevation and improved drainage at the base. Real pocket gophers have spatially distinct, nest-centered territories \citep{lacey2000}. 

\subsection*{Acceptance tests}
\begin{figure}[l]
  {\includegraphics[width=0.4\textwidth]{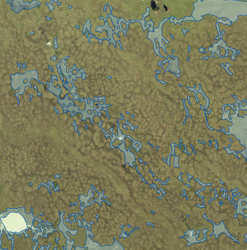}}
  \caption{Aerial photo of a quarter section (0.79 m$^2$) of Burke Ranch, CA, in the wet season. Vernal pools are digitally outlined. Photo from Westervelt Ecological Services.}
\label{fig:Burke}
\end{figure}

\begin{figure}[l]
      \subfigure{\includegraphics[width=0.4\textwidth]{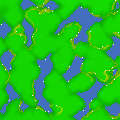}} \qquad
      \subfigure{\includegraphics[width=0.3\textwidth]{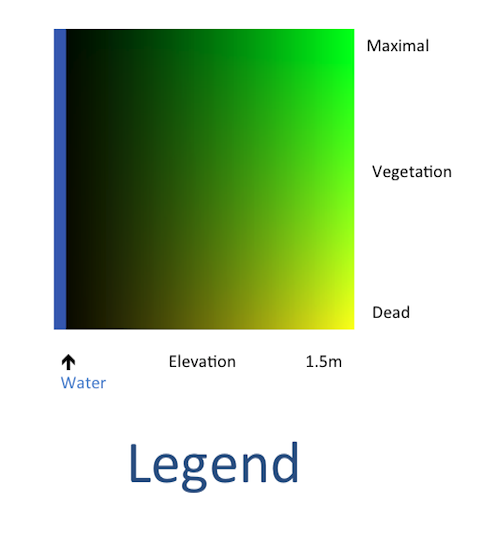}}
    \caption{Results of the full model in 300 simulated years from the beginning of surface inundation. Shown at the highest water level of the year. 12 m a side. } 
  \label{fig:Full}
  \end{figure}

There is aerial photography of a section of the Greater Jepson Prairie during the inundated period available as a real-world example of the degree and kind of organization reported in the real world as Mima mounds (Fig. \ref{fig:Burke}). \cite{keeley1998} characterize mound-pool microtopography as being rain-filled and evapotranspiration-emptied, not connected to overland drainage systems. This distinguishes them from vernal swales and most other temporary wetlands. In the model, we distinguish `pool' from `nonpool' landscapes by checking for this disconnection. An image of the landscape is generated at the most inundated season of the year, and we count distinct underwater regions. This can be done automatically with \textit{ndimage.watershed\_ift}. A landscape averaging at least 0.05 pools/m$^2$ is counted as a `pool' landscape. Most simulations using the final model have a higher density of pools than that (Fig. \ref{fig:Full}). 

Similarly, a `mound' landscape has at least 0.05 mounds/m$^2$, where a mound is a distinct region in the highest 20\% of the distribution of elevations. This distinguishes a mound-pool landscape from a landscape with pools that seem to be pockmarked into a level or branching surface. 

\section*{Results}
\label{sec:results}

In computer simulations, these small interactions are enough to turn a randomly irregular surface into moundlike microtopography without any additional organizing principle  (Fig. \ref{fig:Full}). Identifiable mound-pool structures develop from random or flat topography in approximately 250 simulated years, where observation estimates mound re-formation at 50 or 80 years \citep{JohnsonGSA2011}. 

The minimal full set of tested interactions that generates mound-pool topography, within five hundred simulated years, is: 

\begin{list}{$\circ$}{}
\item Plants grow fastest when neither water\-limited nor underwater.
\item Fossorial rodents move towards the densest adjacent plant growth and
  consume some of it, as in \cite{reichman1985impact}. 
\item Rodents don't tunnel underwater.
\item Rodents will only occupy a cell next to another rodent to avoid drowning.
\item Rodents move a  volume of earth behind them as they travel; maximum volume estimated from  \cite{andersen1981population}.
\item Soil is eroded at intervals during the rainy season. Unvegetated soil erodes more quickly than vegetated soil.
\item The soil is periodically over-saturated. This could represent
  flooding in a drainage basin or seasonal rain in a region with
  impermeable soil layers.
\item The highest water level increases over the first two hundred years. 
\end{list}

The result of these processes is a four-stage feedback loop: 

\begin{list}{$\circ$}{}
\item Irregular topography directs water into depressions
\item Plants grow most richly at the edge of the water
\item In the wet season, rodents live upslope but move downslope to forage
\item Rodents push earth uphill when they move downhill, reinforcing the irregular topography.
\end{list}

The alternate algorithms for each of these that do not produce mound-pool topography, from level, random, or sloped initial topography, are marked in Table \ref{tab:Processes} with `\textit{Fails}'.

There are two parameters to which the system is particularly sensitive. The rodents' mutual avoidance radius may be 0, 1, 2, or 3 cells wide.  Mound-generation improves (from 2/3 success on random initial topography, to success on all initial topography) when the radius is greater than 0 cells wide. Mound generation may occur more quickly with greater radii, but success does not change. The full model uses a 1-cell radius, as the least restrictive.

\begin{figure}[l]
     \includegraphics[width=0.4\textwidth]{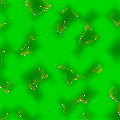}
    \caption{Full model \textit{except} that vegetated soil erosivity is 0.8 that of \texttt{gopher}-excavated tailings; in full model, 0.2. Simulation 450 years from a random initial surface and the initiation of perched water.} 
\label{fig:erosion}
\end{figure}

The difference in erosion rates between vegetated and unvegetated soil is also significant. The full model assumes that vegetated soil erodes 0.2 times as quickly as unvegetated soil, an estimate derived from preliminary observations at Jepson Prairie. If vegetated soil erodes 0.8 times as quickly as unvegetated soil, the landscape develops neither mounds nor pools, as colluvium from the high regions fills the potential pools (Fig. \ref{fig:erosion}). 

\subsection*{Alignment on slopes}
\begin{figure}[l]
  \includegraphics[width=0.4\textwidth]{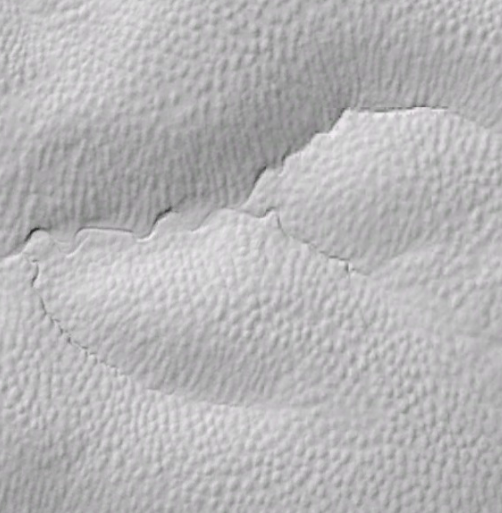}
  \caption{NCALM LiDAR of Mima mounds near Merced, California. Lat 37.433684$^\circ$, lon -~120.418560$^\circ$. Available from http://calm.geo.berkeley.edu/zip/merced\_ug727\_4145.zip, and as a Google Earth overlay.}
\label{fig:MercedSlope}
\end{figure}

\begin{figure}[l]
      \subfigure{\includegraphics[width=0.4\textwidth]{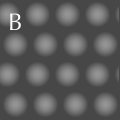}} \qquad
      \subfigure{\includegraphics[width=0.4\textwidth]{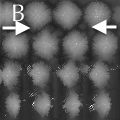}}
    \caption{Elevation plot (light color is elevated) of mounds on a slight slope; aligned (A) or not (B). Slope begins between the first and second row of mounds (between arrows) and moves toward the bottom of the image.  Overland flow moves unvegetated sediment. Erosion is increased below intermound gaps, forcing alignment downslope. Rodent activity moves soil into the protected region below mounds, generating mound extension. Highly stylized hydrology. 12 m a side, simulated 150 years. } 
  \label{fig:Align}
  \end{figure}

One addition to the full model above was a rough test of the system on a slope, adding overland flow and advection of sediment. Terrace landforms on the eastern edge of the Central Valley have well-developed, densely packed Mima mounds (Fig. \ref{fig:MercedSlope}). The mounds extend into the gullies and channels at the edge of the terraces, and on those slopes the mounds are aligned and often extended along the line of fall. In a very rough simulation of overland flow (Fig. \ref{fig:Align}), an initial condition of smooth, divided mounds, aligned or offset, erode into alignment along the fall-line. Rodent activity fills in the space between upslope and downslope mounds, creating extended mounds in the observed direction. 

\begin{figure}[l]
      \subfigure{\includegraphics[width=0.4\textwidth]{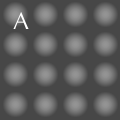}} \qquad
      \subfigure{\includegraphics[width=0.4\textwidth]{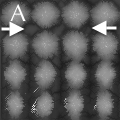}}
\end{figure}

\section*{Discussion}
\label{sec:discussion} 

Emergent mound-pool development, even in simulation, makes the fossorial rodent hypothesis stronger because it explains the co-occurrence of mound-pool systems over restrictive layers or in floodplains without requiring adaptive rodent behavior or initial topographic conditions that we cannot observe or test.  The small-scale interactions it depends on can be observed in the field and adjusted for parametrization or spatial correlation, or ruled out. 

The mound-pool organization occurs because the full model is a reaction-diffusion system, qualitatively like the Belousov-Zhabotinsky reaction, which generates mazes or dots of chemical states depending on ratios of reaction rates. It is easy to get patterns from a reaction-diffusion system \citep{cantrell2003}, so this is a valuable explanation only insofar as the processes match ecosystem processes observed in the field. 



Further research, therefore, will concentrate on fieldwork to examine the spatial relations between the processes in the full model, and in modeling more complex processes that may be significant over geological time.  For instance, the current model ignores rodent nest and territory maintenance. Since these rodents maintain nests and territories in all their habitats, nests are not a sufficient explanatory feature. However, territory geometry might make the mound shapes and sizes more realistic. Rodent attraction to the edge of surface water could be based on a preference for soil of a particular saturation, as in  \citep{andersen1981population}.

\subsection*{Acknowledgements}
Thanks to Ronald Amundson and Sarah E. Reed for discussions of Mima mounds and the fossorial hypothesis, and to John Harte for discussions of ecosystem modeling. 

\bibliographystyle{abbrvnat}
\bibliography{../../Everything}

\end{document}